\documentclass[journal=nalefd,manuscript=article,layout=onecolumn]{achemso}
\setkeys{acs}{articletitle=false,etalmode=truncate,maxauthors=10,keywords=true}

\usepackage{graphicx,color}
\graphicspath{{figs/}}

\author{Jin-Wu Jiang}
\affiliation{Shanghai Institute of Applied Mathematics and Mechanics, Shanghai Key Laboratory of Mechanics in Energy Engineering, Shanghai University, Shanghai 200072, People's Republic of China}
\email{jiangjinwu@shu.edu.cn}

\author{Harold S. Park}
\affiliation{Department of Mechanical Engineering, Boston University, Boston, Massachusetts 02215, USA}

\title{Negative Poisson's Ratio in Single-Layer Graphene Ribbons}

\keywords{Graphene, Negative Poisson's Ratio, Edge Effect, Warping Effect}

\begin{document}

\begin{abstract}

The Poisson's ratio characterizes the resultant strain in the lateral direction for a material under longitudinal deformation.  Though negative Poisson's ratios (NPR) are theoretically possible within continuum elasticity, they are most frequently observed in engineered materials and structures, as they are not intrinsic to many materials.  In this work, we report NPR in single-layer graphene ribbons, which results from the compressive edge stress induced warping of the edges. The effect is robust, as the NPR is observed for graphene ribbons with widths smaller than about 10 nm, and for tensile strains smaller than about 0.5\%, with NPR values reaching as large as -1.51.  The NPR is explained analytically using an inclined plate model, which is able to predict the Poisson's ratio for graphene sheets of arbitrary size.  The inclined plate model demonstrates that the NPR is governed by the interplay between the width (a bulk property), and the warping amplitude of the edge (an edge property), which eventually yields a phase diagram determining the sign of the Poisson's ratio as a function of the graphene geometry.

\end{abstract}

%\pacs{61.48.Gh}
% 61.48.Gh, Structure of graphene
%\keywords{Graphene, Negative Poisson's Ratio, Edge Effect}
%\maketitle
\pagebreak

%\section{Introduction}

The Poisson's ratio ($\nu$) characterizes the resultant strain in the lateral direction for a material under longitudinal deformation.  Most materials contract (expand) laterally when they are stretched (compressed), so that the Poisson's ratio is positive for these materials.  A negative Poisson's ratio (NPR) is allowed by classical elasticity theory, which sets a range of $-1< \nu < 0.5$ for the Poisson's ratio for isotropic materials.\cite{LandauLD}

The NPR has been found to exist intrinsically in some materials, and various models have been proposed for the explanation.\cite{RothenburgL1991nat,LakesR1993adm,EvansKE2000adm,LethbridgeZAD2010am} Milstein and Huang reported NPR for some face centered cubic structures, in which the Poisson's ratio is calculated using the elastic modulus.\cite{MilsteinF1979prb} Baughman et al. explored the correlation between the NPR and the work function in many face centered cubic metals, and the NPR was interpreted from a structural point of view.\cite{BaughmanRH1998nat}  In contrast to individual materials, many more examples of NPR phenomena have been reported in composites and other engineered structures since the seminal work of Lakes in 1987.\cite{LakesR1987sci} In this experiment, the NPR was induced by a permanent compression of a conventional low-density open cell polymer foam, which was explained by the re-entrant configuration of the cell. Materials with NPR have become known as auxetic, as coined by Evans.\cite{EvansKE1991Endeavour}

The study of NPR phenomena has focused on bulk, engineered auxetic structures.\cite{LakesR1993adm,EvansKE2000adm,YangW2004jmsci,GreavesGN2011nm}  However, in the past 3 years reports of NPR phenomena have emerged for low-dimensional nanomaterials.  For example, the NPR for metal nanoplates was found due to a surface-induced phase transformation.\cite{HoDT2014nc}  NPR was found to be intrinsic to single-layer black phosphorus due to its puckered configuration, which leads to NPR in the out-of-plane direction.\cite{JiangJW2014bpnpr} NPR was also predicted for few-layer orthorhombic arsenic using first-principles calculations.\cite{HanJ2015ape} For graphene, we are aware of one report of NPR by Grima et al., which occurred due to the introduction of many vacancy defects, and the resulting rippling curvature, in bulk graphene sheets.\cite{GrimaJN2015adm}

One of the key defining physical characteristics of nanomaterials is their large ratio of surface area to volume (for 1D nanomaterials like nanowires), or edge length to area (for 2D nanomaterials like graphene).  Because of this surface, or edge effects can play a fundamental role in impacting the mechanical properties of these nanomaterials.  For example, as mentioned above surface stress-induced phase transformations were the mechanism enabling NPR in metal nanoplates.\cite{HoDT2014nc}  As graphene is a 2D nanomaterial with the thinnest possible (one atom thick) thickness, its physical properties are very sensitive to free edge effects, for example edge warping due to compressive edge stresses.\cite{ShenoyVB} The warped free edges are also the origin for localized edge phonon modes that are responsible for the edge reconstruction of graphene\cite{JiaX2009sci,EngelundM2010prl} or edge induced energy dissipation in graphene nanoresonators.\cite{KimSY2009nl,JiangJW2012jap}

In this letter, we report, using molecular statics simulations, an intrinsic NPR induced by the warped free edges in single-layer graphene ribbons. The effect is robust, as NPR as large as -1.51 is observed for graphene ribbons with widths smaller than about 10 nm, and for tensile strains smaller than about 0.5\%.  The NPR is explained analytically using an inclined plate model, which is able to predict the Poisson's ratio for graphene sheets of arbitrary size.  The inclined plate model demonstrates that the NPR is governed by the interplay between the width (a bulk property), and the warping amplitude of the edge (an edge property), which eventually yields a phase diagram determining the sign of the Poisson's ratio as a function of the graphene geometry.

\begin{figure}[htpb]
  \begin{center}
    \scalebox{1.0}[1.0]{\includegraphics[width=8cm]{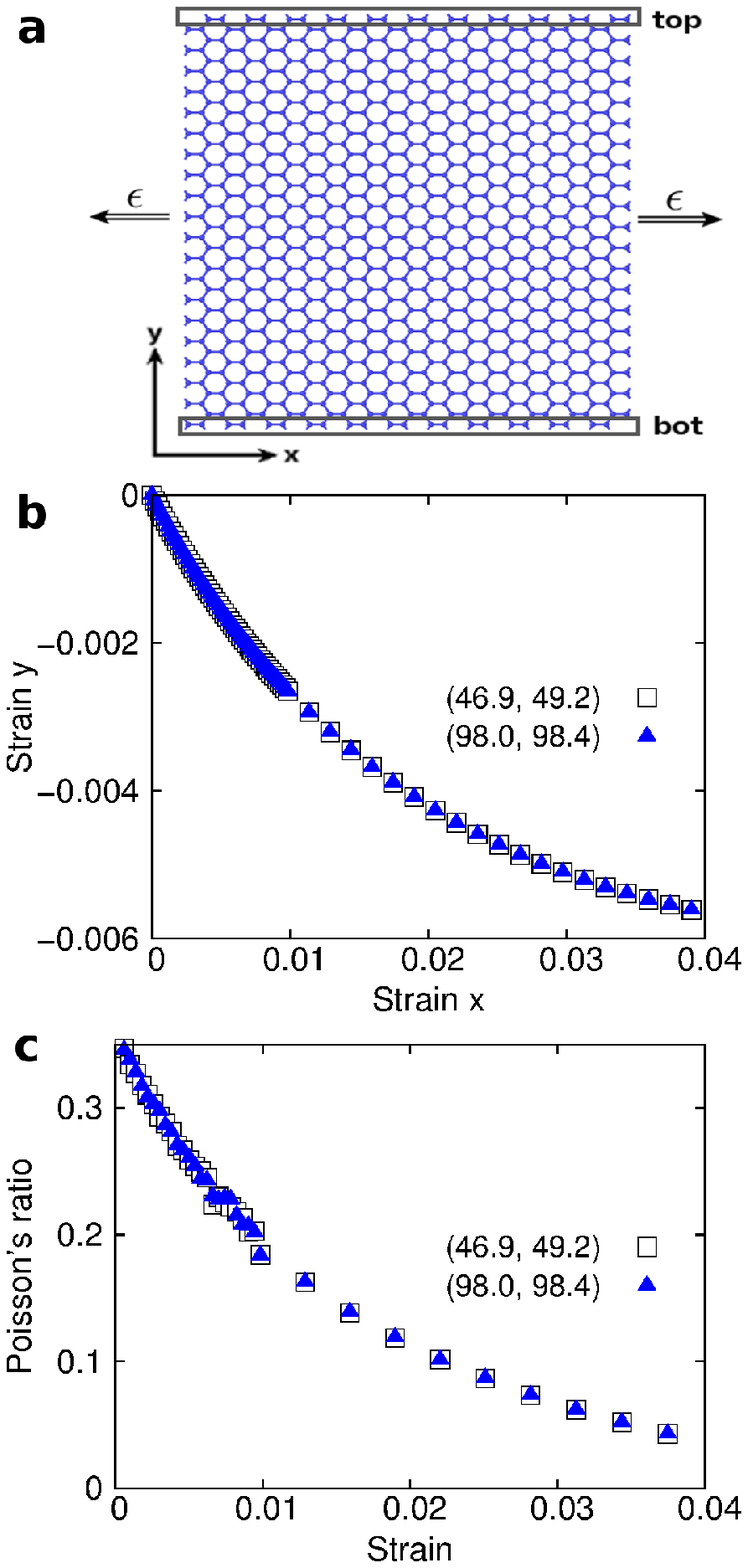}}
  \end{center}
  \caption{(Color online) Poisson's ratio for bulk graphene, i.e. with periodic boundary conditions in both the x and y-directions. a) Graphene is stretched in the x-direction by tensile strain $\epsilon$. The width in the y-direction is described by $W=y_{\rm top}-y_{\rm bot}$, in which $y_{\rm top}$ ($y_{\rm bot}$) is the averaged y coordinate of atoms in the top (bottom) region (in gray boxes). b) The resultant strain in the y-direction versus the applied strain in the x-direction. c) The Poisson's ratio extracted from the strain-strain relation in b.}
  \label{fig_pbc}
\end{figure}

%\section{Poisson's ratio for bulk graphene}
%\section*{Results}
\textbf{Results.} We start by first briefly characterizing the Poisson's ratio in bulk graphene.  To do so, we stretch graphene with periodic boundary conditions (PBC) in both the x and y-directions. Graphene is stretched in the x-direction as shown in Fig.~\ref{fig_pbc}a and the resultant strain in the y-direction is recorded. Fig.~\ref{fig_pbc}b shows the $\epsilon_y$-$\epsilon_x$ relation for graphene of dimensions $46.86\times 49.19$~{\AA} and $97.98\times 98.38$~{\AA}, while Fig.~\ref{fig_pbc}c shows the corresponding Poisson's ratio, where a negligible difference in results from these two differently sized sheets is observed. Fig.~\ref{fig_pbc}c shows that Poisson's ratio in graphene decreases with applied tensile strain, which agrees with the findings of previous continuum mechanics\cite{ReddyCD2006nano} and first-principles calculations.\cite{LiuF2007prb} In the small strain region, the Poisson's ratio value is 0.34, which matches values previously found using the Brenner potential.\cite{ReddyCD2006nano} The salient point is that the Poisson's ratio of graphene is always positive if PBCs, which eliminate edge warping effects, are applied in the y-direction, or the direction normal to the stretching direction.

\begin{figure}[htpb]
  \begin{center}
    \scalebox{1.0}[1.0]{\includegraphics[width=8cm]{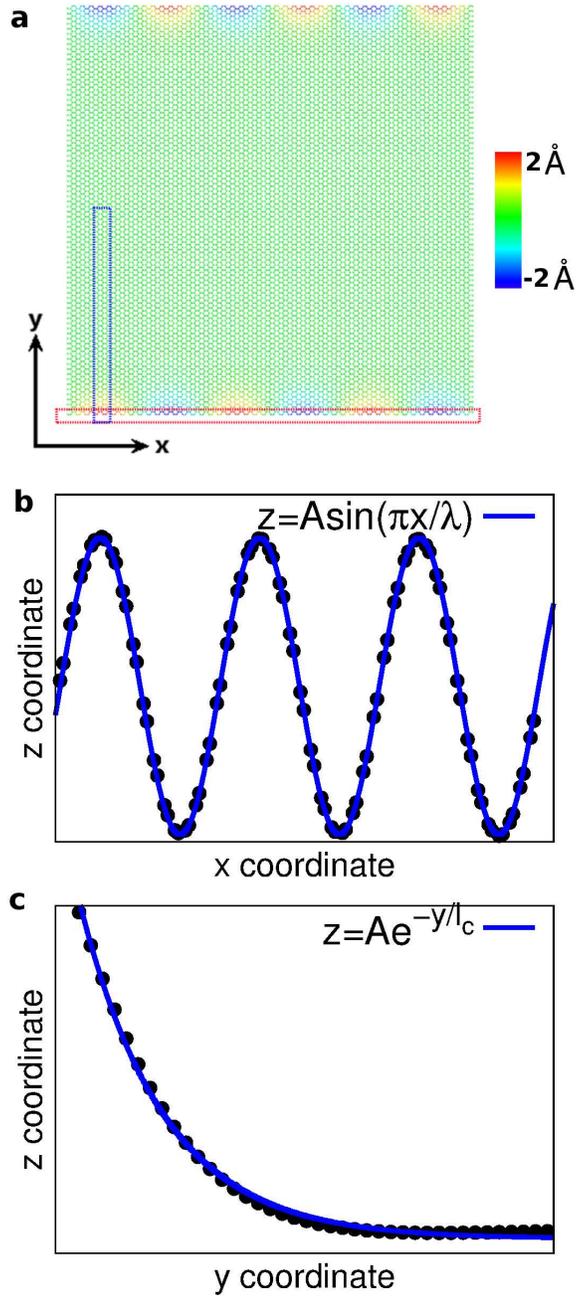}}
  \end{center}
  \caption{(Color online) Warped free edges in graphene. Warping surface is described by $z(x,y)=Ae^{-y/l_c}\sin(\pi x/\lambda)$, with $A=2.26$~{\AA}, $l_c=8.55$~{\AA}, and $\lambda=32.01$~{\AA}. a) Top view of graphene with free boundary conditions in the y-direction, while periodic boundary conditions are applied in the x-direction. Color is with respect to the atomic z-coordinate. b) The z-coordinates for atoms in the horizontal red box. c) The z-coordinates for atoms in the vertical blue box.}
  \label{fig_cfg_fbc}
\end{figure}

%\section{Free Edge-Induced Negative Poisson's Ratio}
A characteristic feature for free edges in graphene is the warped configuration that is induced by the compressive edge stress as shown in Fig.~\ref{fig_cfg_fbc}a. The warped structure can be described by the surface function\cite{ShenoyVB} $z(x,y)=Ae^{-y/l_c}\sin(\pi x/\lambda)$, where $\lambda=L/n$, with $L$ being the length of the graphene ribbon and $n$ being the warping number.  The graphene ribbon shown in Fig.~\ref{fig_cfg_fbc}a has dimensions of $195.96\times 199.22$~{\AA}, resulting in the fitting parameters for the warped free edges as the warping amplitude $A=2.26$~{\AA}, penetration depth $l_c=8.55$~{\AA}, and half wave length $\lambda=32.01$~{\AA}. We note that $\lambda$ is about one sixth of the length $L$, i.e., $\lambda=L/6$.

We study five sets of graphene structures with free boundary conditions in the y-direction. Set I: graphene is 195.96~{\AA} in length, and the warped edge has a warping number $n=6$. Set II: graphene is 195.96~{\AA} in length, and the warped edge has a warping number $n=8$. Set III: graphene is 195.96~{\AA} in length, and the warped edge has a warping number $n=10$. Set IV: graphene is 195.96~{\AA} in length, and the warped edge has a warping number $n=12$. Set V: graphene is 97.98~{\AA} in length, and the warped edge has a warping number $n=2$. For each simulation set, we consider eight different widths of 29.51, 39.35, 49.19, 59.03, 78.70, 98.38, 147.57, and 199.22~{\AA}.  We will demonstrate that the NPR phenomena we report is robust, and is observed for different warping periodicities.

\begin{figure}[htpb]
  \begin{center}
    \scalebox{0.8}[0.8]{\includegraphics[width=8cm]{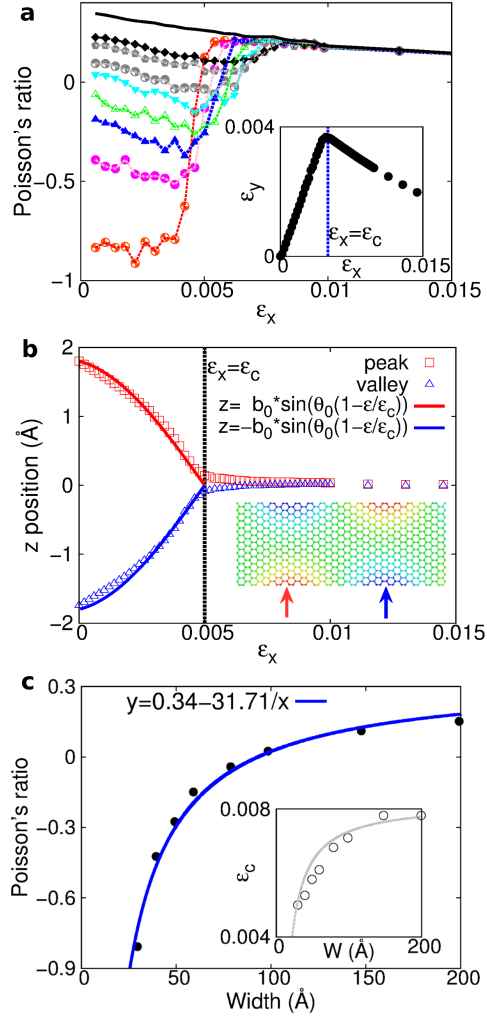}}
  \end{center}
  \caption{(Color online) Poisson's ratio for graphene from Set I. a) Strain dependence for Poisson's ratio. Inset displays the strain-strain relation for graphene with width 29.51~{\AA}, in which graphene expands in the y-direction when it is stretched in the x-direction by strain smaller than $\epsilon_c=0.005$, indicating the NPR effect. b) The strain dependence for the z positions of two atoms in the peak (valley) of the warped edge in graphene. Data are fitted to functions $z=\pm b_0\sin(\theta_0(1-\epsilon/\epsilon_c))$ with constraint $b_0=z_0/\sin\theta_0$. Inset displays these two atoms; i.e., the atom on the peak (red arrow) and valley (blue arrow) of the warped edge. Both atoms fall into the graphene plane for strain larger than $\epsilon_c$. c) Width dependence for the Poisson's ratio. Inset shows the critical strain in graphene ribbons of different width.}
  \label{fig_fbc}
\end{figure}

Fig.~\ref{fig_fbc}a shows the strain dependence for the Poisson's ratio of Set I, where the width of the graphene ribbon increases for data from the bottom to the top in the figure. The occurrence of a NPR is clearly observed for small strains, and for the narrower width ribbons. Furthermore, the Poisson's ratio changes from negative to positive at some critical strain $\epsilon_c$. This critical strain is more clearly illustrated in the inset, which shows a critical strain of $\epsilon_c=0.005$ in the $\epsilon_y$-$\epsilon_x$ relation for graphene with width 29.51~{\AA}. For $\epsilon_x<\epsilon_c$, graphene expands in the y-direction when it is stretched in the x-direction; i.e., the NPR phenomenon occurs.

The critical strain $\epsilon_c$ represents a structural transition for the warped edge. To reveal this structural transition, we show in Fig.~\ref{fig_fbc}b the z position of two atoms from different warped edge regions. One atom is at the peak of the warped edge (shown by the red arrow in the inset), while the other atom is at the valley of the warped edge (shown by blue arrow).  Fig.~\ref{fig_fbc}b clearly shows that both atoms fall into the xy plane at the critical strain $\epsilon_x=\epsilon_c$.  In other words, the warped edge transitions at the critical point from a three-dimensional, out-of-plane warping configuration into a two-dimensional planar configuration due to the externally applied tensile strain. The connection of the critical strain in the disappearance of the NPR in Fig.~\ref{fig_fbc}a and the transition to the two-dimensional planar configuration in Fig.~\ref{fig_fbc}b implies that the NPR is connected to the flattening of the warped edges, as the Poisson's ratio becomes positive after the structural transition of the warped edge. We note that the z-coordinates of the atoms in the warped edges in Fig.~\ref{fig_fbc}b can be well fitted to the functions $z=\pm b_0\sin(\theta_0(1-\epsilon/\epsilon_c))$  for $\epsilon<\epsilon_c$.

The width dependence for the Poisson's ratio is displayed in Fig.~\ref{fig_fbc}c. The Poisson's ratio is strain dependent as shown in Fig.~\ref{fig_fbc}a, so we compute an averaged Poisson's ratio using data in the strain range $[0, \epsilon_c]$, which is equivalent to extracting the Poisson's ratio value by a linear fitting for the $\epsilon_y$-$\epsilon_x$ relation in $[0, \epsilon_c]$. Fig.~\ref{fig_fbc}c shows this averaged Poisson's ratio value for graphene with different widths. We note that the critical strain varies for graphene with different width as indicated by the inset in Fig.~\ref{fig_fbc}c, where the critical strain is fitted to the function $\epsilon_c=0.0082-0.092/W$. The critical strain is smaller in narrower graphene, because some interactions occur between the free ($\pm$y) edges for narrower widths. These edge interactions enable the tension-induced structural transition of the warped edges for narrower ribbons to occur at lower values of tensile strains, because the warping directions for the free ($\pm$y) edges are different, forming a see-saw like configuration. The saturation value $\epsilon_c=0.0082$ at $W\rightarrow\infty$ can be regarded as the actual value of the critical strain for an isolated warped free edge.

The Poisson's ratio in Fig.~\ref{fig_fbc}c increases with increasing width ($W$), and can be fitted to the function $\nu=0.34-31.71/W$. According to this result, graphene can be regarded as the integration of one central region (with bulk Poisson's ratio $\nu_0$) and two edge regions (with Poisson's ratio $\nu_e$). The size of each edge region is $l_c$, which is the penetration depth of the warped configuration in Fig.~\ref{fig_cfg_fbc}. The size of the remaining central region is $W-2l_c$. Simple algebra gives the effective Poisson's ratio for the graphene ribbon as
\begin{eqnarray}
\nu = \nu_{0}-\frac{2l_{c}}{W}\left(\nu_{0}-\nu_{e}\right).
\label{eq_nu}
\end{eqnarray}
Comparing equation~(\ref{eq_nu}) with the fitting function in Fig.~\ref{fig_fbc}c, we get $\nu_0=0.34$ and $\nu_e=-1.51$.

This result shows that for the Set I ribbon geometries, in the limit of an ultra-narrow, edge-dominated graphene ribbon, the value of the NPR can be as large as -1.51. Perhaps more importantly, according to equation~(\ref{eq_nu}), the NPR phenomenon can be observed in graphene sheets with widths up to about 10 nm.  Such width nanoribbons are not small, and are regularly studied experimentally.\cite{JiaoL2009nat}  Furthermore, Fig.~\ref{fig_fbc}a shows that the NPR phenomenon is most significant for tensile strains smaller than about 0.5\%.  These strain values are important as they fall within the strain range of [0, 0.8\%] that has already been achievable in many experimental strain engineering investigations.\cite{NiZH2008acsn,MohiuddinTMG2009prb} We thus expect that this NPR phenomenon can readily be observed experimentally in the near future.

\begin{figure*}[htpb]
  \begin{center}
    \scalebox{1.0}[1.0]{\includegraphics[width=\textwidth]{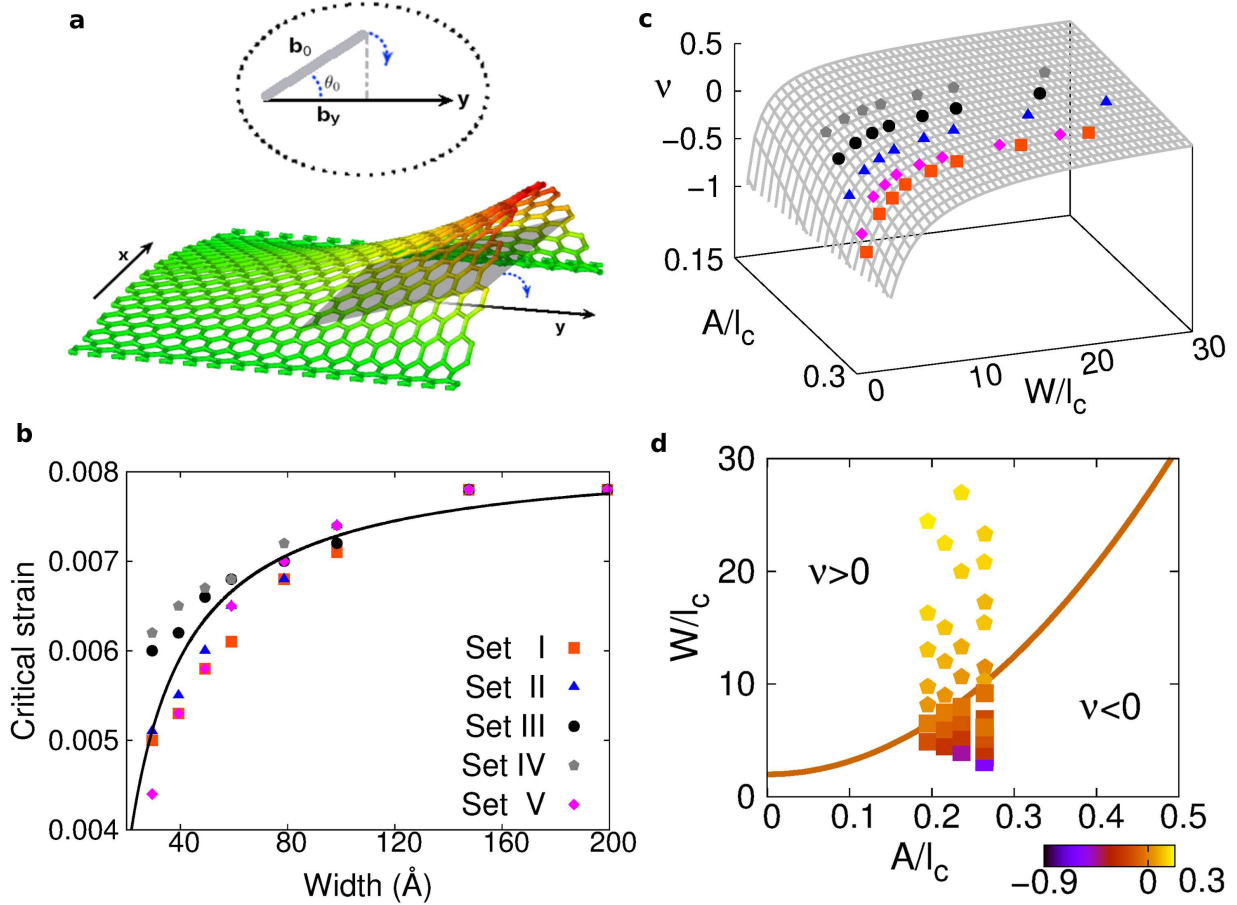}}
  \end{center}
  \caption{(Color online) IP model for warped edge induced NPR. a) The warped free edge is represented by the IP (in gray). During the tensile deformation of graphene, the IP falls down, which leads to the increase of its projection along the y-direction, resulting in the NPR effect. b) Critical strain versus width for all of the five simulation Sets. The solid line is the fitting function $\epsilon_c=0.0082-0.092/W$. c) Comparison of analytic IP model for Poisson's ratio as function of 
 graphene width $\tilde{W}=W/l_c$ and warping amplitude $\tilde{A}=A/l_c$ according to equation~(\ref{eq_nu_ip_general}) for simulation sets I, II, III, IV and V.  d) Phase diagram for positive or negative Poisson's ratio in the $\tilde{W}$ and $\tilde{A}$ parameter space, according to equation~(\ref{eq_diagram}), using all data from all of the five simulation Sets. Color bar is for the value of the Poisson's ratio. }
  \label{fig_ipmodel}
\end{figure*}

%\section{Inclined Plate model for warped edge induced NPR}
%\section*{Discussion}
\textbf{Discussion.} We have demonstrated in the above discussion the connection between the NPR phenomenon and the warped free edges in graphene. We now present an analytic model to describe the relationship between the NPR phenomenon and the warped free edges. In Fig.~\ref{fig_ipmodel}a, the warped edge is represented by an inclined plate (IP) (gray area). During the tensile deformation, the IP falls into the xy plane. The side view in the dashed ellipse illustrates the mechanism enabling the NPR clearly.  Specifically, it shows that the projection ($b_y$) of the IP on the y-axis increases during the falling down of the IP, resulting in the NPR phenomenon.

This IP model is inspired by the strain-dependent z (out-of-plane) coordinates of atoms in the warped edges shown in Fig.~\ref{fig_fbc}b, where the z-coordinates are fitted to the functions $z=\pm b_0\sin(\theta_0(1-\epsilon/\epsilon_c))$, in which the parameters are restricted by $b_0=z_0/\sin\theta_0$. This function describes exactly the trajectory of the tip of the IP (displayed by blue arrow in Fig.~\ref{fig_ipmodel}a) during its falling down process in Fig.~\ref{fig_fbc}b. This function also indicates that the IP's tilting angle $\theta$ is a linear function of the applied tensile strain $\epsilon$,
\begin{eqnarray}
\theta = \theta_{0}\left(1-\frac{\epsilon}{\epsilon_{c}}\right),
\label{eq_theta}
\end{eqnarray}
where $\theta_0$ is the initial tilting angle. This expression gives $\theta=0$ at the critical strain $\epsilon=\epsilon_{c}$, as required by the definition of the critical strain in Fig.~\ref{fig_fbc}b. For the applied tensile strain $\epsilon$ in the x-direction, the resulting strain in the y-direction is
\begin{eqnarray*}
\epsilon_{y} = \frac{\cos\theta-\cos\theta_{0}}{\cos\theta_{0}}\approx\frac{\theta_{0}^{2}}{\epsilon_{c}}\epsilon,
\end{eqnarray*}
yielding the Poisson's ratio of the edge
\begin{eqnarray}
\nu_{e} & = & -\frac{\epsilon_{y}}{\epsilon_{x}}=-\frac{\theta_{0}^{2}}{\epsilon_{c}}.
\label{eq_nu_ip}
\end{eqnarray}

We now determine the initial tilting angle $\theta_{0}$ for the IP. The tilting angle with respect to the y-axis for the tangent plane at point $\left(x,y,z\right)$ on the warped surface is
\begin{eqnarray*}
\phi\left(x,y\right) & = & \tan^{-1}\left(\frac{\partial w}{\partial y}\right)\approx\frac{A}{l_{c}}e^{-y/l_{c}}\sin\frac{\pi x}{\lambda},
\end{eqnarray*}
in which the tilting angle is assumed to be small. This assumption is reasonable as will be shown below. The average tilting angle for the warping area $x\in[0,\lambda]$ and $y\in[0,l_{c}]$ is
\begin{eqnarray}
\bar{\phi} & = & \frac{1}{\lambda l_{c}}\int_{0}^{\lambda}dx\int_{0}^{l_{c}}dy\phi\left(x,y\right)=\frac{A}{l_{c}}\frac{2}{\pi}\left(1-\frac{1}{e}\right).
\label{eq_phi}
\end{eqnarray}
Inserting the value of $A$ and $l_{c}$ from Fig.~\ref{fig_cfg_fbc}, we get an average tilting angle $\bar{\phi}=0.106$. We use this average tilting angle as the initial tilting angle for the IP, i.e., $\theta_{0}=\bar{\phi}=0.106$. As a result, we obtain the Poisson's ratio for the warped edge
\begin{eqnarray*} 
\nu_{e}=-\frac{\theta_{0}^{2}}{\epsilon_{c}}=-\left(\frac{A}{l_{c}}\frac{2}{\pi}\left(1-\frac{1}{e}\right)\right)^{2}/\epsilon_{c}=-1.37
\end{eqnarray*}
in which the critical strain $\epsilon_{c}=0.0082$ is the saturation value from the inset of Fig.~\ref{fig_fbc}c, which should be used here in the discussion of the Poisson's ratio for an isolated warped edge. The Poisson's ratio of -1.37 for the Set I geometries that is obtained in the limit of an ultra-narrow graphene ribbon using the IP model is very close to the value of -1.51 obtained via the molecular statics calculations in Fig.~\ref{fig_fbc}c.

We find in Fig.~\ref{fig_ipmodel}b that the critical strains for Sets I-V obey the same relation $\epsilon_c=0.0082-0.092/W$, where again we use the saturation value of $\epsilon_{c}=0.0082$ for the calculation of the effective Poisson's ratio of graphene ribbons with warped edges. The critical strain is directly related to the compressive edge stress, which generates compressive strain ($\epsilon_{e}$) in the edge region, and thus various warped configurations (with different local minimum potential energies).  From equation (\ref{eq_theta}), the magnitude of the compressive edge strains equals the critical strain $\epsilon_{c}$ when the warped edges have been completely flattened into a planar structure, i.e. $\epsilon_{e}=\epsilon_{c}=0.0082$.  Using the calculated values for the Young's modulus of the edge region as $E_{e}=11.15$8~{eV\AA$^{-2}$}, while the compressive edge stress density is\cite{ShenoyVB} $\sigma_{e}=1.05$~{eV\AA$^{-1}$}, we can  estimate the width ($W_e$) of the region that will be compressed by the edge strain according to $W_e=\sigma_{e}/\left(E_{e}\epsilon_e\right)=11.48$~{\AA}. Taking a representative value of the penetration depth $l_{c}=8.55$~\AA~from simulation Set I, we find that the penetration depth of the resultant warped configuration is close to $W_{e}$. These analysis illustrate that the saturation value of the critical strain is an intrinsic property for the free edge, which should not depend on the detailed warping configuration.

From equations~(\ref{eq_nu}), ~(\ref{eq_nu_ip}) and ~(\ref{eq_phi}), we can thus obtain the following general formula for the Poisson's ratio in graphene ribbons of arbitrary width,
\begin{eqnarray}
\nu = \nu_{0}-\frac{2}{\tilde{W}}\left(\nu_{0}+\frac{1}{\epsilon_{c}}\tilde{A}^{2}C_{0}^{2}\right),
\label{eq_nu_ip_general}
\end{eqnarray}
where $C_{0}=\frac{2}{\pi}\left(1-\frac{1}{e}\right)$ is a universal constant and $\nu_{0}=0.34$ is the Poisson's ratio for bulk graphene. The dimensionless quantity $\tilde{W}=W/l_{c}$ is the width with reference to the penetration depth $l_{c}$. This is a bulk related quantity, and a larger $\tilde{W}$ correlates with moving the Poisson's ratio in the positive direction. The other dimensionless quantity $\tilde{A}=A/l_{c}$ is the warping amplitude with reference to the penetration depth. This quantity is an edge related property, which tunes the Poisson's ratio in the negative direction.  The sign and also the value of the Poisson's ratio are determined by the competition between these two effects, while the length of graphene has no effect. Fig.~\ref{fig_ipmodel}c shows a three-dimensional plot for the Poisson's ratio predicted by the IP model based on equation~(\ref{eq_nu_ip_general}). The numerical results for all of the five simulation Sets are also shown in the figure, which agree quite well with the IP model.

We note that the effective Poisson's ratio defined in equation~(\ref{eq_nu_ip_general}) is intrinsically width-dependent, due to the width-dependence of the warping amplitude $A$ and the penetration depth $l_{c}$.  Thus, by using the dimensionless quantities $\tilde{W}$ and $\tilde{A}$, the effect of the warping number is intrinsically included in the expression for the effective Poisson's ratio.  Furthermore, the resultant equation~(\ref{eq_nu_ip_general}) is a general expression for the Poisson's ratio, which is an explicit function of the ribbon geometry. Hence, equation~(\ref{eq_nu_ip_general}) could be readily extended to describe the Poisson's ratio in other similar atomic-thick materials.

By equating equation~(\ref{eq_nu_ip_general}) to be zero, we get
\begin{eqnarray}
\tilde{W} = 2+\frac{2}{\nu_{0}}\frac{C_{0}^{2}}{\epsilon_{c}}\tilde{A}^{2}.
\label{eq_diagram}
\end{eqnarray}
This function is plotted in Fig.~\ref{fig_ipmodel}d, and serves to delineate the positive and negative Poisson's ratio regions. This figure serves as a phase diagram for NPR phenomenon in the parameter space of $\tilde{A}$ and $\tilde{W}$. The NPR phenomenon occurs for graphene with parameters in the region below the curve. In particular, if the width $\tilde{W}<2$, then the Poisson's ratio is negative irrespective of the value for the other parameter $\tilde{A}$. The numerical results for all of the five simulation Sets can be correctly categorized into positive or negative regions in this phase diagram, further validating the analytical IP model we have presented.

%\section{Conclusion}
\textbf{Conclusion.} We have used molecular statics simulations to demonstrate the occurrence of negative Poisson's ratios in graphene ribbons.  The negative Poisson's ratios occur due to the warping of the free edges due to the compressive edge stresses, and can be observed for graphene ribbons with widths smaller than about 10 nm, and for tensile strains smaller than about 0.5\%.  An inclined plate model was developed, which revealed the link between the warped free edge, and the negative Poisson's ratios.  The model also led to an analytic formula for the structural dependence for the Poisson's ratio of graphene.  Specifically, we find that the value of the Poisson's ratio is determined by the interplay between the width (bulk property) and the warping amplitude (edge property), while the length does not play a role. The analytic phase diagram for the sign and value of the Poisson's ratio accurately explained  the numerical results. We expect these results to further augment the already interesting suite of physical properties and potential applications for graphene ribbons.\cite{JiaX2009sci,EngelundM2010prl}

%\section{Methods}
%\section*{Methods}
\textbf{Methods.} The NPR studies of single-layer graphene ribbons were performed using classical molecular statics simulations, where the interactions between carbon atoms in graphene were described by the second generation Brenner potential~\cite{brennerJPCM2002}, which has been widely used to study the mechanical response of graphene.\cite{MoY2009nat} The Cartesian coordinate system is displayed in Fig.~\ref{fig_pbc}a.  Periodic boundary conditions (PBCs) were applied in the x-direction, while either PBCs or free boundary condition (FBCs) were applied in the y-direction.

To obtain stable freestanding graphene ribbons with edge warping, following Shenoy et al.\cite{ShenoyVB}, we introduce a small out-of-plane perturbation $0.5e^{-y/10.0}\sin(\pi x/\lambda)$ to the z-coordinate of edge atoms in the ideal planar graphene structure.  This perturbed structure is then relaxed by conjugate gradient (CG) energy minimization.  The relaxation is performed until the relative energy change is less than $10^{-12}$.  We note that the half wave length $\lambda=L/n$, with the warping number $n$ being an even number corresponding to the PBCs applied in the x-direction and $L$ being the length of the graphene ribbon. In other words, if a perturbation with $\lambda=L/6$ is used, then the relaxed structure shown in Fig.~\ref{fig_cfg_fbc}a is obtained. If the perturbation with $\lambda=L/8$ is used, then the relaxed warped edge has $\lambda=L/8$.  The effect of the warping number $n$ is that if $n$ is larger, there are more warping segments at the edge, though each segment has a smaller warping amplitude.  

We find that, for the graphene ribbons of length 195.96~{\AA}, the warped configurations with warping number $n\geq 14$ are not energetically favorable, as they transition into warped configurations with smaller warping number during tensile deformation.  On the other hand, the energy difference among these warped configurations with warping number $n\leq 4$ is very small, so the graphene ribbons will assume multiple warping configurations during tensile deformation.  We thus have focused our investigations into configurations with warping number $6\leq n \leq 12$.

Once the equilibrium warped configuration is obtained, the tensile loading is applied.  This is done by stretching the relaxed graphene ribbon in the x-direction as illustrated in Fig.~\ref{fig_pbc}a by deforming the simulation box in the x-direction. The deformed structure is then relaxed by the CG energy minimization procedure, in which the structure also allowed to be fully relaxed in the lateral directions. All simulations were performed using the publicly available simulation code LAMMPS~\cite{PlimptonSJ,Lammps}. The OVITO package was used for visualization~\cite{ovito}.

Corresponding to the applied external tension $\epsilon_x=\epsilon$ in the x-direction, the resultant strain in the y-direction is computed as
\begin{eqnarray}
\epsilon_{y} & = & \frac{W-W_{0}}{W_{0}},
\end{eqnarray}
where the width is $W=y_{\rm top}-y_{\rm bot}$ with $y_{\rm top}$ ($y_{\rm bot}$) being the averaged y-coordinate for atoms from the top (bottom) group in Fig.~\ref{fig_pbc}a. The Poisson's ratio is then calculated by its definition
\begin{eqnarray}
\nu = -\frac{\epsilon_{y}}{\epsilon_x}.
\label{eq_nu_def}
\end{eqnarray}
In our numerical calculation, equation~(\ref{eq_nu_def}) is realized using the finite difference method. More specifically, the Poisson's ratio at strain $\epsilon_x=\epsilon^j$ is obtained by
\begin{eqnarray}
\nu_j = -\frac{\epsilon_{y}^{j+1}-\epsilon_{y}^{j-1}}{\epsilon^{j+1}-\epsilon^{j-1}},
\end{eqnarray}
where the integer j=1, 2, ..., represents the strain increment number.

\textbf{Acknowledgements} The work is supported by the Recruitment Program of Global Youth Experts of China, the National Natural Science Foundation of China (NSFC) under Grant No. 11504225 and the start-up funding from Shanghai University. HSP acknowledges the support of the Mechanical Engineering department at Boston University.

\textbf{Author contributions} J.W.J performed the calculations and discussed the results with H.S.P. J.W.J and H.S.P co-wrote the paper.

\textbf{Competing financial interests} The authors declare no competing financial interests.

%\bibliographystyle{aipnum4-1}
%\bibliographystyle{nature}
%\bibliography{biball}
%\bibliography{/home/JiangJinWu/Documents/papers/mypapers/latex/biball}

\providecommand{\latin}[1]{#1}
\providecommand*\mcitethebibliography{\thebibliography}
\csname @ifundefined\endcsname{endmcitethebibliography}
  {\let\endmcitethebibliography\endthebibliography}{}

\end{document}